\documentclass[aps,prl,twocolumn,groupedaddress]{revtex4}
\usepackage[dvips]{graphicx}
\usepackage{hyperref}
\usepackage{amssymb}

\begin{document}

\title{Mapping photonic entanglement into and out of a quantum memory}
\author{K. S. Choi}
\thanks{These authors contributed equally to this work.}
\affiliation{Norman Bridge Laboratory of Physics 12-33, California Institute of
Technology, Pasadena, California 91125, USA}
\author{H. Deng}
\thanks{These authors contributed equally to this work.}
\affiliation{Norman Bridge Laboratory of Physics 12-33, California Institute of
Technology, Pasadena, California 91125, USA}
\author{J. Laurat\footnote{Present Address : Laboratoire Kastler Brossel, Universit\'{e} Paris 6, Ecole
Normale Superieure et CNRS, UPMC Case 74, 4 place Jussieu, 75252 Paris Cedex 05, France}}
\affiliation{Norman Bridge Laboratory of Physics 12-33, California Institute of
Technology, Pasadena, California 91125, USA}
\author{H. J. Kimble}
\email[To whom correspondence should be addressed. E-mail : ]{hjkimble@caltech.edu}
\affiliation{Norman Bridge Laboratory of Physics 12-33, California Institute of
Technology, Pasadena, California 91125, USA}
\date{\today}
\maketitle

\textbf{Recent developments of quantum information science \cite{zoller05} critically rely on \textit{entanglement}, an intriguing
aspect of quantum mechanics where parts of a composite system can exhibit
correlations stronger than any classical counterpart \cite{clauser78}.
In particular, scalable quantum networks require capabilities to create,
store, and distribute entanglement among distant matter nodes via photonic
channels \cite{briegel00}. Atomic ensembles can play the role of such nodes
\cite{duan01}. So far, in the photon counting regime, heralded
entanglement between atomic ensembles has been successfully demonstrated via
probabilistic protocols \cite{chou05,chou07}. However, an inherent drawback
of this approach is the compromise between the amount of entanglement and
its preparation probability, leading intrinsically to low count rate for
high entanglement. Here we report a protocol where entanglement between two
atomic ensembles is created by coherent mapping of an entangled state of
light. By splitting a single-photon \cite{tan91,hessmo04,Jacques07} and subsequent state transfer,
we separate the generation of entanglement and its storage \cite{Sangouard07}. After a
programmable delay, the stored entanglement is mapped back into photonic
modes with overall efficiency of $17\%$. Improvements of single-photon
sources \cite{Lounis05} together with our protocol
will enable \textquotedblleft on-demand\textquotedblright\ entanglement of
atomic ensembles, a powerful resource for quantum networking.}
\vspace{0.4cm}

In the quest to achieve quantum networks over long distances \cite{briegel00},
an area of considerable activity has been the
interaction of light with atomic ensembles comprised of a large collection
of identical atoms \cite{duan01,review3,giacobino}. In the regime of
continuous variables, a particularly notable advance has been the
teleportation of quantum states between light and matter \cite{polzik}. For
discrete variables with photons taken one by one, important achievements
include the efficient mapping of collective atomic excitations to single
photons \cite{laurat06,felinto06,thompson06,Kuzmich06,chen06}, the
realization of entanglement between a pair of distant ensembles \cite{chou05,laurat07b}
and, more recently, entanglement distribution involving
two pairs of ensembles \cite{chou07}. The first step toward entanglement
swapping has been made \cite{laurat07a} and light-matter teleportation has
been demonstrated \cite{pan07}.

In all these cases, progress has relied upon probabilistic schemes
following the measurement-induced approach developed in the seminal paper by
Duan, Lukin, Cirac and Zoller \cite{duan01} (\textit{DLCZ}) and subsequent
extensions. For the \textit{DLCZ} protocol, heralded entanglement is
generated by detecting a single photon emitted indistinguishably by
one of two ensembles. Intrinsically, the probability $p$ to prepare
entanglement with only $1$ excitation shared between two ensembles is
related to the quality of entanglement, since the likelihood for
contamination of the entangled state by processes involving $2$ excitations scales as $p$ \cite{laurat07b},
and results in low success probability for each trial. Although the degree of
stored entanglement can approach unity for the (rare) successful trials \cite{laurat07b},
the condition $p\ll1$ dictates reductions in count rate and
compromises in the quality of the resulting entangled state (e.g., as $p\rightarrow0$,
processes such as stray light scattering and detector dark
counts become increasingly important). Furthermore, for finite memory time,
subsequent connection of entanglement becomes increasingly challenging \cite{laurat07a}.

\begin{figure*}[tbph]
\includegraphics[width=1.8\columnwidth]{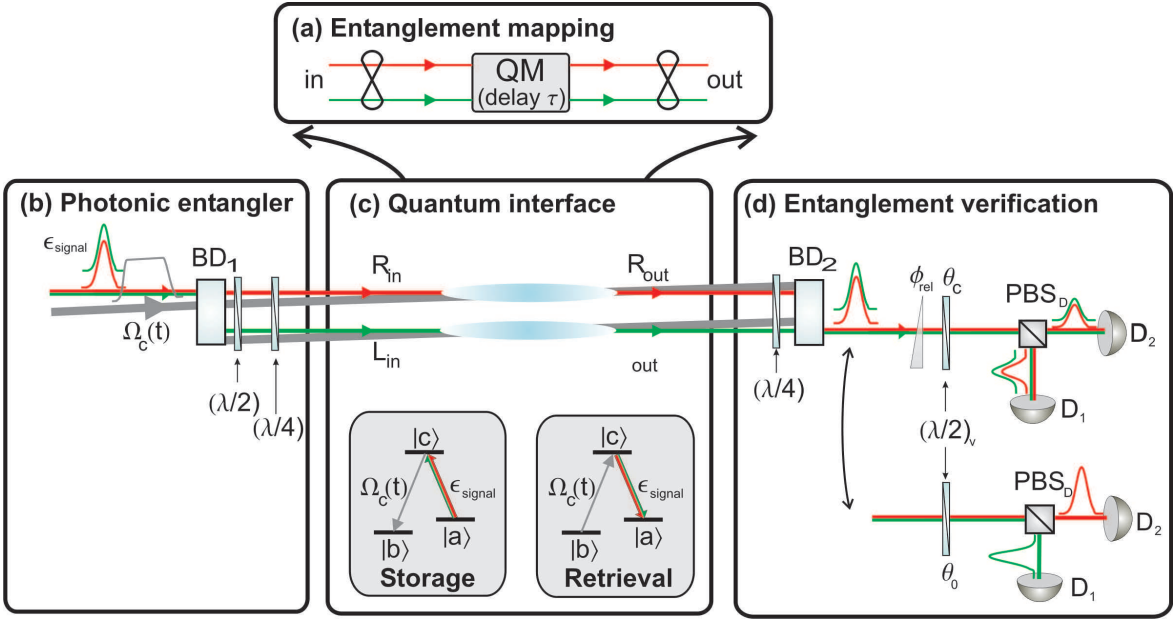}
\caption{\textbf{Overview of the experiment.} \textbf{a,} Illustration of
the mapping of an entangled state of light into and out of a quantum memory
(QM) with storage time $\protect\tau$. \textbf{b,} Photonic
\textquotedblleft entangler\textquotedblright : A beam displacer $BD_{1}$
splits an input single photon into two orthogonally polarized, entangled
modes $L_{in},R_{in}$, which are spatially separated by 1 mm. With
waveplates $\protect\lambda/2$ and $\protect\lambda/4$, the signal fields $\epsilon_{signal}$
for $L_{in},R_{in}$ and control fields $\Omega_{c}^{(L,R)}(t)$ are
transformed to circular polarizations with the same helicity along each path $L,R$, and copropagate with angle of $3^{\circ}$.
\textbf{c,} Reversible mapping : Photonic entanglement between $L_{in},R_{in}$ is
coherently mapped into the memory ensembles $L_{a},R_{a}$
by switching $\Omega_{c}^{(L,R)}(t)$ off adiabatically. After a
programmable storage time, the atomic entanglement is reversibly mapped back
into optical modes $L_{out},R_{out}$ by switching $\Omega_{c}^{(L,R)}(t)$ on.
Relevant energy diagrams for the storage and retrieval processes are shown
in the insets. States $|a\rangle$, $|b\rangle$ are the hyperfine ground
states $F=4$, $F=3$ of $6S_{1/2}$ in atomic cesium; state $|c\rangle$ is
the hyperfine level $F^{\prime }=4$ of the electronic excited state $6P_{3/2}$.
\textbf{d,} Entanglement verification : After a $\protect\lambda/4$ plate, the beam
displacer $BD_{2}$ combines modes $L_{out},R_{out}$ into one beam with
orthogonal polarizations. With $(\protect\lambda /2)_{v}$ at $\theta_{c}=22.5^{\circ}$ before the
polarization beamsplitter (PBS$_{D}$), single photon interference is recorded at detectors
$D_{1},D_{2}$ by varying the relative phase $\protect\phi_{rel}$ by a Berek compensator. With $(\protect\lambda/2)_{v}$
at $\theta_{0}=0^{\circ}$, photon statistics for each mode $L_{out},R_{out}$ are
measured independently.}
\label{setup}
\end{figure*}

The separation of processes for the generation of entanglement and for its
storage enables this drawback to be overcome. Here, we demonstrate such a
division by way of reversible mapping of an entangled state into a
quantum memory. The mapping is obtained by using adiabatic passage based
upon dynamic Electromagnetically Induced Transparency (EIT) \cite{Fleischhauer00,kash99,Harris97,Hau99}
(See Appendix). Storage and retrieval of a
single photon have been demonstrated previously \cite{Chaneliere05,Eisaman05}. Adiabatic
transfer of a collective excitation has been demonstrated between two
ensembles coupled by a cavity mode \cite{vuletic07}, which can provide a
suitable approach for generating on-demand entanglement over short
distances. However, for efficient distribution of entanglement over quantum
networks, reversible mapping of an entangled state between matter and light,
as illustrated in Fig. \ref{setup}a, has not been addressed until now.

In our experiment, entanglement between two atomic ensembles $L_{a},R_{a}$
is created by first splitting a single photon into two modes $L_{in},R_{in}$
to generate an entangled state of light \cite{tan91,Jacques07,hessmo04}. This entangled
field state is then coherently mapped to an entangled matter state for $L_{a},R_{a}$.
On demand, the stored atomic entanglement for $L_{a},R_{a}$ is reversibly converted back into entangled
photonic modes $L_{out},R_{out}$. As opposed to the original \textit{DLCZ}
scheme, our approach is inherently deterministic, suffering principally from
the finite efficiency with which single excitations can be mapped to and
from an atomic memory, which can approach $45\%$ \cite{Novikova07}. Moreover,
the contamination of entanglement for the $L_{a},R_{a}$ ensembles from
processes involving $2$ excitations can be arbitrarily suppressed
(independent of the mapping probabilities) with continuing advances
in on-demand single photon sources \cite{Lounis05}.
Our experiment thereby provides a
promising avenue to distribute and store entanglement deterministically
over remote atomic ensembles for quantum networks \cite{Sangouard07}.

The experimental setup is depicted in Fig. \ref{setup}. Our single photon
source is based on Raman transitions in an optically thick cesium ensemble
\cite{duan01,laurat06} (See Appendix). This system generates 25 ns-long single
photons (resonant with the $6S_{1/2},F=4\leftrightarrow6P_{3/2},F^{\prime}=4$ transition) in a heralded
fashion \cite{laurat06}. The single photons are polarized at 45$^{\circ}$
from the eigen-polarizations of the beam displacer $BD_{1}$ (Fig. 1b), which
splits them into entangled optical modes $L_{in},R_{in}$ (called the signal modes) to produce, in
the ideal case, the state $\frac{1}{\sqrt{2}}(|0_{L_{in}}\rangle
|1_{R_{in}}\rangle +e^{i\phi_{rel}}|1_{L_{in}}\rangle |0_{R_{in}}\rangle )$.

The next stage consists in coherently mapping photonic entanglement for
$L_{in},R_{in}$ into atomic ensembles $L_{a},R_{a}$ (called the memory ensembles)
within a single cloud of cold cesium atoms in a magneto-optical trap (MOT) (Fig. 1c).
Ensembles $L_{a},R_{a}$ are defined by the well-separated optical paths of
the entangled photonic modes $L_{in},R_{in}$. To avoid dissipative
absorption for the fields in modes ${L_{in},R_{in}}$ for our choice of
polarization \cite{Chaneliere05}, we initially spin-polarize the atomic
ensemble into $|F=4,m_{F}=0\rangle$ (See Appendix). A synchronous clock
governs the trials of both the single photon source and memory ensembles
with a period of 575 ns. Initially, the strong control fields $\Omega_{c}^{(L,R)}$ (resonant
with the $6S_{1/2},F=3\leftrightarrow6P_{3/2},F^{\prime}=4$ transition) open
transparency windows $\Omega_{c}^{(L,R)}(0)$ in $L_{a},R_{a}$ for the
signal modes. As the wavepacket of the signal field propagates through each
ensemble, the control fields $\Omega_{c}^{(L,R)}(t)$ are turned off in 20 ns
by an intensity modulator, thus coherently transforming the fields of
the respective signal modes to collective atomic excitations within $L_{a},R_{a}$.
This mapping leads to heralded entanglement between ensembles $L_{a},R_{a}$.
After a user-defined delay, chosen here to be 1.1 $\mu$s, the
atomic entanglement is converted back into entangled photonic modes by
switching on the control fields $\Omega_{c}^{(L,R)}(t)$ (See Appendix).

For a given optical depth $\gamma$, there is an optimal Rabi frequency $\Omega_{c}(t)$
for the control field. In our experiment, the measured $\gamma$ and $\Omega_{c}(0)$
are $15$ and $24$ MHz, respectively. An
example of our measurements of the EIT process for a single ensemble is
presented in Fig. \ref{srcoherent}, which shows in (a) the input
single-photon pulse and in (b) its storage and retrieval (See Appendix). Due
to finite $\gamma$ and small length ($\approx3$ mm) of
the ensemble, there is considerable loss in the storage process, as
evidenced by the counts around $\tau=0$ $\mu$s in (b). The peak beyond $\tau=1$ $\mu$s
represents the retrieved pulse after $1.1$ $\mu$s of storage. Overall, we find
good agreement between our measurements and the results from a numerical
calculation following the methods of \cite{Fleischhauer00}, using the
fitted function of the input signal field (Fig. \ref{srcoherent}a) as the
initial condition with all other parameters from independent measurements.
The overall storage and retrieval efficiency is measured to be $\eta_{r}=17\pm1\%$,
also in agreement with the simulation.

\begin{figure}[tbph]
\includegraphics[width=0.9\columnwidth]{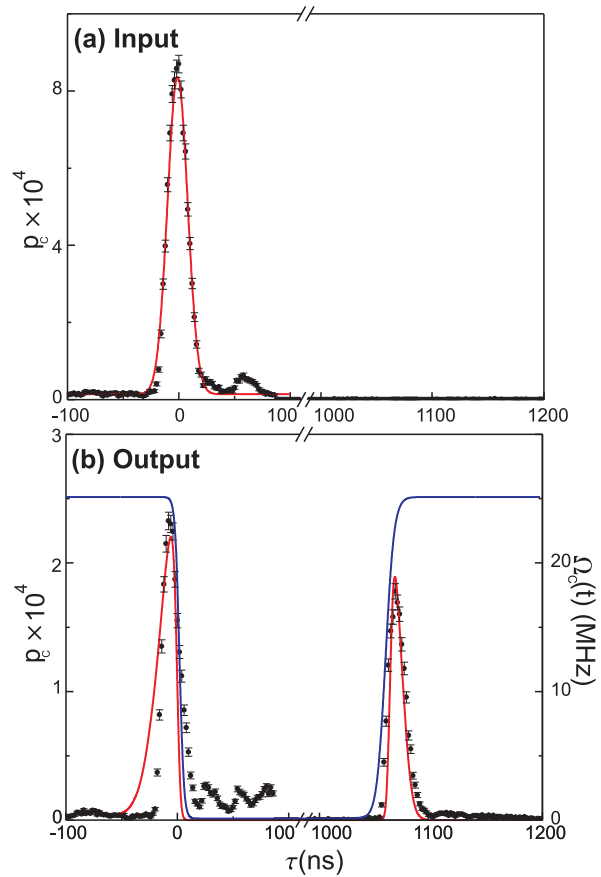}
\caption{\textbf{Single photon storage and retrieval for a single ensemble.}
\textbf{a,} Input : The measured probability density $p_{c}$ for signal
field, here a single photon generated from a separate atomic ensemble
\protect\cite{laurat06}. The red solid line represents a gaussian fit of
$1/e$ width of $28$ ns.
\textbf{b,} Storage and retrieval : The points around $\protect\tau=0$ $\protect\mu$s
represent \textquotedblleft leakage\textquotedblright\ of the signal field
due to the finite optical depth and length of the ensemble. The points
beyond $\protect\tau=1$ $\protect\mu$s show the retrieved signal field.
The overall storage and retrieval efficiency is $17\pm1\%$. The blue solid
line is the estimated Rabi frequency $\Omega_{c}(t)$ of the control pulse.
The red solid curve is from a numerical calculation solving the equation of
motion of the signal field in a dressed medium \cite{Fleischhauer00}.
Error bars give the statistical error for each point.}
\label{srcoherent}
\end{figure}

With these results in hand for the individual $L_{a},R_{a}$ ensembles, we next turn
to the question of verification of entanglement for the input $L_{in},R_{in}$
and output $L_{out},R_{out}$ optical modes. We follow the protocol introduced in
Ref. \cite{chou05} by (1) reconstructing a reduced density matrix $\rho$
constrained to a subspace containing no more than one
excitation in each mode, and (2) assuming that all off-diagonal elements between
states with different numbers of photons vanish, thereby obtaining a lower
bound for any purported entanglement. In the photon-number basis $|n_{L},m_{R}\rangle$
with $\{n,m\}=\{0,1\}$, the reduced density matrix $\rho$ is written as \cite{chou05}
\begin{equation}
\rho =\frac{1}{P}\left(
\begin{array}{cccc}
p_{00} & 0 & 0 & 0 \\
0 & p_{01} & d & 0 \\
0 & d^{\ast } & p_{10} & 0 \\
0 & 0 & 0 & p_{11}%
\end{array}%
\right) .  \label{rho}
\end{equation}
Here, $p_{ij}$ is the probability to find $i$ photons in mode $L_{k}$ and $j$
in mode $R_{k}$, $d\simeq\frac{V(p_{01}+p_{10})}{2}$ is the coherence
between $|1_{L}0_{R}\rangle_{k}$ and $|0_{L}1_{R}\rangle _{k}$,
$P=p_{00}+p_{01}+p_{10}+p_{11}$, and $V$ is the visibility for
interference between modes $L_{k},R_{k}$, with $k\in{\{in,out\}}$. The
degree of entanglement of $\rho$ can be quantified in terms of the
concurrence $C=\frac{1}{P}max(0,2|d|-2\sqrt{p_{00}p_{11}})$
which is a monotone function of entanglement, ranging from $0$ for a
separable state to $1$ for a maximally entangled state \cite{wooters98}.

Operationally, the various elements of $\rho$ are obtained by recombining
the $L_{k},R_{k}$ fields with a second beam displacer, $BD_{2}$, as illustrated in
Fig. \ref{setup}d, to obtain a single spatial mode with orthogonal
polarizations for the $L_{k},R_{k}$ fields \cite{chou07,laurat07b}. The diagonal
elements of $\rho$ are measured with $(\lambda/2)_{v}$ set at $0^{\circ}$
so that detection events at $D_{1},D_{2}$ are recorded directly for the $L_{k},R_{k}$
fields. To determine the off-diagonal components of $\rho$, the modes $L_{k},R_{k}$
are brought into interference with $(\lambda/2)_{v}$ set at $22.5^{\circ}$,
as shown in Fig. \ref{setup}d. By varying the relative phase
$\phi_{rel}$ between the modes, we determine the visibility for
single-photon interference and thereby deduce $d$.

\begin{figure*}[t!]
\includegraphics[width=1.8\columnwidth]{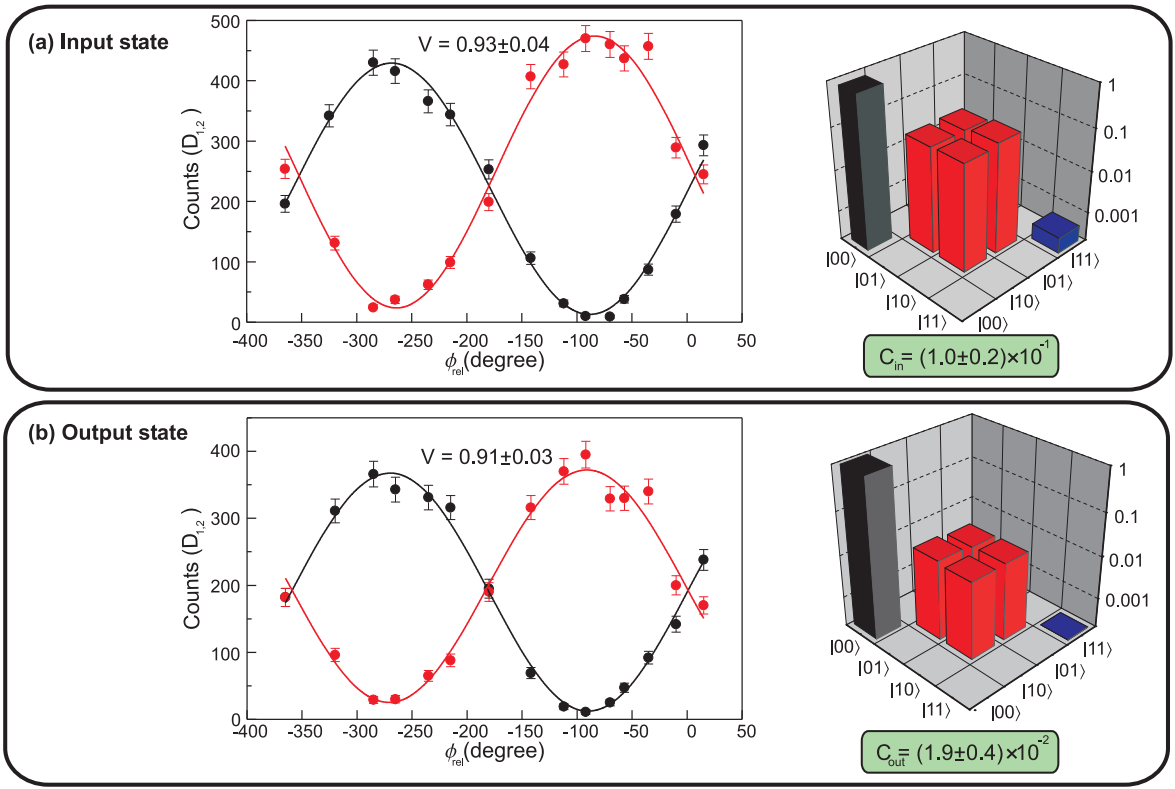}
\caption{\textbf{Entanglement for the input (a) and output (b) optical modes.}
To verify entanglement, complementary measurements are performed :
interference leading to a fringe when the relative phase $\protect\phi_{rel}$
is scanned and photon statistics for the light modes separately. The figure shows the interference fringes
and the reconstructed density matrices (in log scale) for the photonic modes \textbf{a,} at the
input of the memory and \textbf{b,} at the output after storage and
retrieval. The estimated concurrence is given in each case. Each point of
the fringe is taken for 20,000 (100,000) heralding events for the input (output) state.
Error bars indicate statistical errors.}
\label{fringe}
\end{figure*}

We first perform tomography on the input modes ${L_{in},R_{in}}$ to verify
that they are indeed entangled. To this end, we remove the memory ensembles
to transmit directly the signal fields into the verification stage, following
our protocol of complementary measurements. The interference fringe between
the two input modes is shown in Fig. \ref{fringe}a. From the independently
determined propagation and detection efficiencies, our measurements at $D_{1},D_{2}$
can be used to infer the quantum state for the input modes ${L_{in},R_{in}}$ entering
the input faces of the atomic ensembles $L_{a},R_{a} $ \cite{chou05}, with the
reconstructed density matrix $\rho_{in}$ also given in Fig. \ref{fringe}a.
The concurrence derived from $\rho_{in}$ is $C_{in}=0.10\pm 0.02$,
so that the fields for ${L_{in},R_{in}}$ are indeed entangled. The value of the concurrence is in good
agreement with the independently derived expectation of $C_{in}^{theory}=0.10\pm 0.01$,
which depends on the quality of the single photon and the vacuum
component (i.e., the overall efficiency) \cite{laurat07b}.
Given a heralding click from our single photon source, the probability to have a
single photon at the face of either memory ensemble is 15$\%$, leading to a vacuum
component of 85$\%$. We also independently characterize the
suppression $w$ of the two-photon component relative to a coherent state (for which $w=1$)
and find $w=0.09\pm0.03$. Our input entanglement is
only limited by the current properties of our single-photon source, which will be
improved with the rapid advances in sources of single photons
\cite{Lounis05}.

Having verified entanglement for the input modes ${L_{in},R_{in}}$, we next
map this photonic entanglement into the $L_{a},R_{a}$ ensembles, which serve
as a quantum memory (Fig. \ref{setup}c). After storing the entanglement for $1.1$ $\mu$s,
we transfer the resulting atomic excitation from the
memory to the output modes ${L_{out},R_{out}}$ and perform quantum-state
tomography to determine $\rho_{out}$ precisely as for $\rho_{in}$. As
shown in Fig. \ref{fringe}, the visibility for interference of the fields
after storage and retrieval shows no appreciable degradation (from $V_{in}=0.93\pm0.04$
to $V_{out}=0.91\pm0.03$). From the measurements at $D_{1},D_{2}$, we infer
the quantum state $\rho_{out}$\ at the output faces of the $L_{a},R_{a}$
ensembles, with the result displayed in Fig. \ref{fringe}b. The associated
concurrence $C_{out}=(1.9\pm 0.4)\times10^{-2}$ is in agreement with
$C_{out}^{theory}=(1.7\pm 0.1)\times10^{-2}$. Since the mapping of the
atomic states from $L_{a},R_{a}$ into field modes $L_{out},R_{out}$ is a
local operation, this measurement provides a lower bound for the
entanglement between the $L_{a},R_{a}$ ensembles \cite{chou05}. Thus, we
demonstrate the reversible mapping of an entangled state of the
electromagnetic field to and from a material system. For completeness,
Table I gives the diagonal elements and concurrences of $\bar{\rho}_{in},\bar{\rho}_{out}$
determined directly at $D_{1},D_{2}$ without correction for propagation and
detection efficiencies.

We emphasize that although the entanglement associated with $\rho_{out}$ is
heralded (because of the nature of our source of single photons), our
protocol for generation and storage of entanglement is intrinsically
deterministic. The transfer efficiency of entanglement from input modes
to output modes of the quantum memory is limited by the storage and retrieval efficiency $\eta_{r}$ of
the EIT process. This transfer can be quantified by the ratio $\lambda$ of
the concurrence $C_{out}$ for the output state $\rho_{out}$ to $C_{in}$ for
the input state $\rho_{in}$. For an ideal source of single photons
on-demand (with no vacuum component), the input concurrence is approximated
by $C_{in}\simeq\alpha V$, where $\alpha $ denotes the transmission
efficiency of the single photon from the source to the entangler in Fig. \ref{setup}b \cite{laurat07b}.
Similarly, for the output, $C_{out}\simeq\alpha\eta_{r}V$,
where we assume that the visibility $V$ is preserved by the
mapping processes. Thus, $\lambda=\frac{C_{out}}{C_{in}}\simeq\eta_{r}$,
which therefore estimates the maximum amount of entanglement in modes $L_{out},R_{out}$
for the case of an (ideal) single photon generated
deterministically. In our experiment, the entanglement transfer reaches $\lambda =(20\pm5)\%$.

The performance of our quantum interface depends also on the memory lifetime
$\tau_{m}$ over which one can faithfully retrieve a stored quantum state.
For our system, independent measurements of $\eta_{r}$ made by varying the storage
duration $\tau$ allow us to determine $\tau_{m}=8\pm1$ $\mu$s, as limited
by inhomogeneous Zeeman broadening \cite{Hugues05} and
misalignment between the quantization axis and the bias magnetic
field (See Appendix). Active and passive compensations of the residual
magnetic field would improve $\tau_{m}$ \cite{ringot01}, along with
improved optical trapping techniques.

\begin{table}[b]
\caption{Experimentally determined diagonal elements $\bar{p}_{ij}$ and
concurrences $\bar{C}_{in},\bar{C}_{out}$ for the density matrices
$\bar{\protect\rho}_{in},\bar{\protect\rho}_{out}$ derived directly from detectors
$D_{1},D_{2}$ without correction for losses and detection efficiencies. The error bars indicate statistical errors.}
{\footnotesize
\begin{tabular*}{3.5 in}{|l@{\extracolsep{\fill}}cc|}
\hline$ $&$\bar{\rho}_{in}$&$\bar{\rho}_{out}$\\
\hline
$\bar{p}_{00}$&$0.9800\pm0.0001$&$0.99625\pm0.00003$\\
$\bar{p}_{10}$&$(1.043\pm0.008)\times 10^{-2}$&$(2.09\pm0.02)\times 10^{-3}$\\
$\bar{p}_{01}$&$(0.957\pm0.008)\times 10^{-2}$&$(1.67\pm0.02)\times 10^{-3}$\\
$\bar{p}_{11}$&$(8\pm2)\times 10^{-6}$&$(2\pm2)\times 10^{-7}$\\
$\bar{C}$&$(1.28\pm0.09)\times 10^{-2}$&$(2.5\pm0.5)\times 10^{-3}$\\
\hline
\end{tabular*}
}
\label{table}

\end{table}

In conclusion, our work provides the first realization of mapping an
entangled state into and out of a quantum memory. Our protocol alleviates
the significant drawback of probabilistic protocols \cite{duan01}, where low
preparation probabilities prevent its potential scalability \cite{laurat07a},
and thus our strategy leads to efficient scaling for high-fidelity quantum
communication \cite{Sangouard07}. Our current results are
limited by the large vacuum component of our available single photon source,
which principally reduces the degree of entanglement in the input, and by
the limited retrieval efficiency of the EIT process, which bounds the
entanglement transfer $\lambda=20\pm5\%$. However, with extension to
an on-demand single photon source and with improved retrieval efficiency,
pushed already to $45\%$ in Ref. \cite{Novikova07} by larger optical depth
and optimum pulse shaping, our protocol
provides an alternate strategy for generating and distributing entanglement
between remote quantum memories that circumvents some difficulties with the
original \textit{DLCZ} protocol \cite{duan01}.
\vspace{1cm}

\section{APPENDIX}

\subsection{I. Experimental details.}
A 22 ms preparation stage and 3 ms experiment run are
conducted every 25 ms period. During the preparation stage, atomic ensembles
are loaded in a MOT for 18 ms and further cooled by optical molasses for 3 ms.
For $800$ $\mu$s, we optically pump the atomic ensembles to the $6S_{1/2},F=4,m_{F}=0$
state in atomic cesium. During this stage, the
trapping beam is turned off while the intensity of the repumping beam is
reduced to $0.1I_{sat}$. The quantization axis is chosen along the $k$-vector of
the signal modes and defined by a pulsed magnetic field of $0.2$ G. A pair of counter-propagating
Zeeman pumping beams ($10$ MHz red-detuned from $4\leftrightarrow 4^{\prime}$ and
linearly polarized along the
quantization axis) illuminate the ensembles in a direction perpendicular to
modes $L_{in},R_{in}$. The MOT repumping beam serves as a hyperfine pumping
beam. The experiment is conducted at repetition rate of $1.7$ MHz during a $3$ ms
interval before the next MOT loading cycle. A small bias field
of $10$ mG is left on to define the quantization axis for the experiment.

\subsection{II. Single photon generation.}
The single-photon source is based upon the
protocol \cite{duan01,laurat06} composed of time-delayed photon pairs,
called fields 1,2 emitted from a cesium ensemble in a MOT called the source
ensemble, located 3 m from the memory ensembles. For photon-pair
production, a sequence of write and read pulses illuminates the source
ensemble. The single photon generation is heralded by probabilistic
detection of a Raman scattered field 1 from a write pulse. Conditioned on
the heralding signal, a strong read pulse maps the excitation into a photonic
mode, field 2, with probability of $50\%$, which then propagates to the
setup described in Fig. \ref{setup}. The resulting conditional probability to
have a single photon, field 2, at the face of memory ensemble is $15\%$. The
heralding signal triggers a control logic which disables the single-photon
source and all associated laser beams for the programmable duration of the
storage process for the quantum interface.

\subsection{III. EIT storage and retrieval.}
The coherent interface between the signal modes
and collective spin waves is achieved by dynamically controlling the EIT
window $\Omega_{c}(t)$, defined by the atom-light interaction of a resonant
control field. A quantum field propagating through an externally controlled
dressed state medium is best described as a slow-light, dark-state polariton
(DSP), $\hat{\Psi}(z,t)$ \cite{Fleischhauer00}, a coherent mixture of
matter-like and photonic excitations, expressed as
\begin{equation}
\hat{\Psi}(z,t)=\cos{\theta(t)}\hat{\varepsilon}_{signal}-\sin{\theta(t)}\sqrt{N}\hat{\sigma}_{ab}  \label{dark-state}
\end{equation}%
where $\cos ^{2}{\theta(t)}=\frac{\Omega_{c}^{2}(t)}{\Omega_{c}^{2}(t)+g^{2}N}=\frac{\upsilon_{g}(t)}{c}$, $N$ is the number of atoms,
$\hat{\sigma}_{ab}$ is the atomic coherence operator for ground states $|a\rangle $,$|b\rangle $,
and $\Omega_{c}(t)$ is the Rabi frequency of
control field. As the signal field propagates through the medium, the group
velocity $\upsilon_{g}$ of the DSP is adiabatically reduced to zero as $\Omega_{c}(t)$
decreases to zero, thereby rotating $\theta(t)$ from a
purely photonic state to a matter-like collective spin coherence.
When the control field is re-activated, the collective spin excitation is
coherently converted into a photonic mode in a time-reversal fashion.

\vspace{1 cm}

\subsection{Acknowledgement}
We gratefully acknowledge our ongoing collaboration with S. J. van Enk. This
research is supported by the Disruptive Technologies Office and by the
National Science Foundation. H.D. acknowledges support as Fellow of the
Center for the Physics of Information at Caltech. J.L. acknowledges
financial support from the European Union (Marie Curie Fellowship).

\end{document}